\documentclass[sigconf,authorversion,nonacm]{acmart}

\acmISBN{978-1-4503-XXXX-X/2018/06}

\begin{document}

\title{Timeliness, Consensus, and  Composition of the Crowd: Community Notes on X}


\author{Olesya Razuvayevskaya}
\affiliation{%
  \institution{The University of Sheffield}
  \city{Sheffield}
  \country{United Kingdom}
}
\email{o.razuvayevskaya@sheffield.ac.uk}

\author{Adel Tayebi}
\affiliation{%
  \institution{CY Cergy Paris University}
  \institution{NOVA University Lisbon}
  \city{Paris}
  \country{France}}
\email{adel.tayebi@cyu.fr}

\author{Ulrikke Dybdal Sørensen}
\affiliation{%
  \institution{Aalborg Universitet}
  \city{Aalborg}
  \country{Denmark}}
  \email{ulrikkeds@ikp.aau.dk}

\author{Kalina Bontcheva}
\affiliation{%
 \institution{The University of Sheffield}
 \city{Sheffield}
 \country{United Kingdom}}
 \email{K.Bontcheva@sheffield.ac.uk}

\author{Richard Rogers}
\affiliation{%
  \institution{University of Amsterdam}
  \city{Amsterdam}
  \country{Netherlands}}
  \email{R.A.Rogers@uva.nl}

\renewcommand{\shortauthors}{Trovato et al.}

\begin{abstract}
This study presents the first large-scale quantitative analysis of the efficiency of X’s Community Notes, a crowdsourced moderation system for identifying and contextualizing potentially misleading content. Drawing on over 1.8 million notes, we examine three key dimensions of crowdsourced moderation: participation inequality, consensus formation, and timeliness. Despite the system’s goal of collective moderation, we find substantial concentration effect, with the top 10\% of contributors producing 58\% of all notes (Gini Coefficient = 0.68). The observed consensus is rare—only 11.5\% of notes reach agreement on publication, while 69\% of posts receive conflicting classifications. A majority of noted posts ($\approx$68\%) are annotated as “Note Not Needed”, reflecting the repurposing of the platform for debate rather than moderation. We found that such posts are paradoxically more likely to yield published notes ($OR = 3.12$). Temporal analyses show that the notes, on average, are published 65.7 hours after the original post, with longer delays significantly reducing the likelihood of consensus. These results portray Community Notes as a stratified, deliberative system dominated by a small contributor elite, marked by persistent dissensus, and constrained by timeliness. We conclude this study by outlining design strategies to promote equity, faster consensus, and epistemic reliability in community-based moderation. 

\end{abstract}




\received{7 October 2025}

\maketitle

\section{Introduction}
Online content moderation of misinformation typically relies on either automated detection or a manual approach, be it expert fact-checking or crowd-sourced moderation. As the volume of false information continues to grow, manual expert fact-checking is becoming an increasingly intractable task, requiring expertise across multiple languages, topics, and region-specific domains \cite{martel2024crowds, stencel2024half}. Moreover, it carries risks of partisanship, bias and censorship, stemming from the judgments of a small group of experts \cite{flamini2019fact}. In an effort to address scalability and bias concerns, social media platforms have increasingly turned to crowdsourcing as an alternative to expert fact-checking. This shift, started by Twitter (now X) with the introduction of Birdwatch in January 2021 to the users in the United States, was later extended worldwide and renamed Community Notes (CNs). This was followed by Meta's announcement\footnote{https://www.nytimes.com/2025/01/07/business/meta-community-notes-x.html} in early 2025 that it would discontinue its fact-checking program and adopt CNs.

The move toward crowd-sourced moderation signals a change in the philosophy of moderation from expert adjudication to one based on consensus building \cite{de2025twitter}, despite the fact that CNs still rely on fact-checking sources, with around 33\% of notes reported to cite at least one such source \cite{borenstein-etal-2025-community}. This change in philosophy has prompted a series of observations concerning potential issues with crowdsourced moderation. One concerns the prospect that it will facilitate biased or politically "asymmetric interventions" against misinformation \cite{mosleh2024differences}. Another is that CNs can be vulnerable to coordinated attacks, potentially suppressing helpful notes by downvoting them \cite{augenstein2025community}. Along those lines, the polarised nature of online communities could make crowdsourcing itself difficult to achieve\cite{wirtschafter2023future}. These concerns raise a central question: How effective are CNs in achieving the goal of crowdsourced content moderation?

To address this question, in this work, we undertake three complementary analyses. First, we examine the composition of the crowd in terms of potential concentration of note production by a select group of users. Second, we assess the extent to which CNs promote consensus or dissensus—both at the level of individual posts and within the crowd-based evaluation of notes themselves. Within this analysis, we also look into the case of the misuse of CNs as a debating platform. Third, we examine certain conditions of consensus-making, particularly the dependence on the timeliness of producing CNs. By addressing these questions, we can evaluate the rate of consensus building in crowdsourced moderation, specific conditions that can affect this goal and whether CNs are "crowdsourced" in practice or function more as a specialised moderation activity dominated by a narrow set of contributors. 

\section{Background}
\subsection{Community Notes on X} \label{related:platform}
Similar to Wikipedia and Reddit’s models of community-driven moderation, the CN system on X relies on a community of users identifying misleading content or adding important context to the shared information, with the primary goal of surfacing high-quality information \cite{Wirtschafter_Majumder_2023,corsi_crowdsourcing_2024}. To become part of the community, known as a \textit{contributor}, there is a set of eligibility criteria to be met, such as a six-month user history and verified phone number to prevent multiple accounts \cite{10.1145/3706599.3716230}. Upon becoming a contributor, a user accrues `influence' by rating notes to eventually earn the privilege of a \textit{note writer}. To maintain writer status, a contributor must ensure that the majority of the notes they produce are rated as helpful by the community.  

Eligible users write notes about the posts they consider misinformation or potentially misleading, with the option to add explanation tags concerning why the post should be `noted'. They can also create notes to mark a post as non-misinformation, which some note writers do to counter another contributor’s misinformation-flagging note. A note is then rated by the community as helpful or not, which is decisive for the note to be displayed. Only helpful notes that flagged misinformation are published under the post on the main platform. A note has two weeks to achieve a status, and within that period a note's status can change at any point. After that, the note's status is ‘locked’.

All the notes are assigned the status ‘needs more ratings’ (NMR) when first created and must be rated at least five times before being considered for display. A note must not only be rated by enough contributors who consider it helpful; the contributors must also represent a ‘diversity of perspectives’. This is X's method of reaching consensus through a so-called bridging algorithm \cite{ovadya_bridging_2023, wojcik_birdwatch_2022} rather than a simple majority of votes. Crucially, X does not incorporate any explicit attributes of contributors when inferring perspectives; ‘diversity’ is derived entirely from statistical patterns in how contributors have rated other notes. This process can be described as computational consensus defined not by direct deliberation or explicit representation, but by algorithmic classification. 

Previous studies have argued that this computational approach to consensus building introduces complications, since coming to an agreement is not only a technical determination, but also a social task \cite{de_keulenaar_content_2025, matamoros-fernandez_importance_2025}. The social aspect of consensus formation became apparent after X added a new rating category in 2023: NotHelpful\_NoteNotNeeded (NNN). CN contributors use this feature to indicate that the post represents an opinion rather than a check-worthy claim, and therefore does not require a note. Despite the presence of this rating feature, some  prefer to create a new note under the post, with the note text mentioning that the note is not needed. This behaviour is typically motivated by the desire to counter the other notes under the post that mark it as potentially misleading. 

\subsection{Crowdsourced content moderation}
Despite the ongoing questions about the effectiveness of community-driven models for content moderation, such models are increasingly adopted by social media platforms in the form of an additional context alongside posts \cite{lloyd_beyond_2025}. The adoption of such models is based on the idea of leveraging the crowd and harnessing its aggregated insights \cite{wojcik_birdwatch_2022}. For the platforms, leveraging the crowd can be a beneficial approach even in noisy and inefficient user environments, as it relies on accumulated accuracy \cite{woolley_evidence_2010}. It is also considered scalable in addressing a large volume of information, with broader geographical coverage and faster detection \cite{augenstein_community_2025, lloyd_beyond_2025}. 

Community-based content moderation is often regarded as a less intrusive alternative to traditional, top-down fact-checking systems. It represents a bottom-up mechanism for monitoring information and shaping collective judgments about content credibility\cite{augenstein_community_2025}. A growing body of literature that examines crowd moderation in comparison to expert fact-checking has recently emerged in relation to CNs. Within this area of research, there is no agreement on whether crowdsourcing can address some of the limitations inherent in expert fact-checking \cite{borenstein_can_2025}. While some studies have shown that crowd moderation can effectively identify low-quality news sources and that community ratings often  align closely with professional fact-checkers' assessments \cite{martel2024crowds},  others found differences in how the crowd and fact-checkers select and evaluate content \cite{saeed_crowdsourced_2022}.

Certain operational realities of crowdsourced approaches have also become evident. These systems are not necessarily working as intended, indicating that efforts to harness the `wisdom of the crowd' may rest on overly optimistic assumptions about the integrity, diversity, and efficacy of user collaboration \cite{augenstein_community_2025}. Applying the community-model to social media rests on the belief that the crowd has the ability to effectively moderate content; however,  this perspective may overlook the difference of being able to moderate and choosing to do so \cite{prollochs_community-based_2022}. Indeed, community-driven approaches arguably need to address the challenges of the willingness of its users to participate, as user contributions may be driven by political agendas or by purposive, alternative interpretations of facts \cite{kahan_misconceptions_2017, m_otala_political_2021}.

\subsection{Effectiveness of Community Notes on X} \label{sec:effectiveness}
There is a growing academic interest in evaluating the effectiveness of community-based moderation approaches, with particular attention to X's CNs, whose open-source data offer an opportunity for empirical analysis \cite{wang_efficiency_2024}. Key research questions concern the relationship between community-based moderation and  professional fact-checking, the impact of CNs on the misleading posts they annotate, the capacity of the community to reach consensus as well as the system's overall capacity to mitigate politically motivated polarization.

With respect to the timeliness of CNs, previous studies have shown that CN moderation frequently lags during the initial and often most viral phase of content dissemination, implying that the system's current speed may not be sufficient to combat the circulation of misleading content \cite{renault_collaboratively_2024}. This observation further highlights concerns about scalability of the system, suggesting that the effectiveness of moderation may not increase proportionally as the user base grows.  

The effectiveness of the system also depends heavily on the quality and characteristics of sources, since CNs frequently draw upon fact-checked materials \cite{borenstein_can_2025,drolsbach_community_2024, prollochs_community-based_2022}. Solovev et al. have found that references to external sources significantly increase the odds of the note being perceived as helpful \cite{solovev_references_2025}. These findings are consistent with prior research on note credibility, especially in the context of Covid-19 vaccines, which reported that notes addressing misleading posts were typically of good quality \cite{allen_characteristics_2024}. Other studies have found that the sources referenced in the notes are highly factual which in turn contributes to greater rater consensus \cite{kangur_who_2024}.

Another research direction focuses on the impact of published notes on content perception and dissemination. Prior studies have found a remarkable reduction in the spread of misleading information as a result of the post being `noted' \cite{chuai_community-based_2024}. Evidence suggests that when misinformation is annotated, the authors of such posts show concern for their reputation, frequently leading to broad retractions \cite{gao_can_2024}. Moreover, posts with notes indicating misleading information have seen a decline in their visibility through a reduction in reposting compared to the posts determined to be not misleading \cite{drolsbach_diffusion_2023}. Notes also affect user engagement, with the number of shares experiencing decrease by approximately 50\%, while replies and views experience smaller, though still noticeable, declines \cite{slaughter_community_2025}.

Consensus formation within the CN system represents a key bottleneck, constraining the overall effectiveness of moderation. Empirical evidence underscores this challenge: only a small fraction of all created notes become visible, and a limited proportion of contributors produce notes that achieve consensus \cite{braga_2025}. Contributors have been found to be more likely to write negative notes on posts by counter-partisans and to rate counter-partisans' notes as unhelpful \cite{allen_birds_2022}. Furthermore, Augenstein et al. highlight contributor behaviour and assessment dynamics as a barrier to the system's effectiveness, including its vulnerability to coordinated behaviour \cite{augenstein2025community, elliott_elon_nodate}. 
Aligned with this observation, some researchers contend that the level of polarization on X is such that the bridging algorithm  appears unable, perhaps even by design, to moderate the most highly polarized content \cite{bouchaud_algorithmic_2025, wirtschafter_future_2023}. Given that the effectiveness of CNs depends on the ability to achieve consensus, failure to bridge the partisan divide can result in misleading content remaining unaddressed \cite{prollochs_community-based_2022,de_keulenaar_content_2025}. 

In an effort to increase the effectiveness of the system, X continually updates and introduces new features. One such feature is the option for users to request notes on posts, where the goal is to increase the system's scale. Chuai et al. found that such requests prompt contributors to prioritise posts they characterise as greatly misleading, concurrently de-prioritising political content \cite{chuai_community_2025}. Another direction of research concerns the use of LLMs to generate ``supernotes", thereby overcoming the prevalent dissensus phenomenon  \cite{de_supernotes_2024, mohammadi_ai_2025}.

As can be seen, prior research on CNs has largely focused on system-level outcomes, such as overall reductions in the spread of misleading content, without systematically investigating how the composition of the contributor crowd or the distribution of note production affects moderation dynamics. Additionally, the role of CNs as a platform for debate, rather than strictly as a moderation tool, has received limited attention. Our study addresses these gaps by combining analyses of contributor behaviour, note-level agreement, and temporal dynamics. This provides a more granular understanding of the mechanisms underlying CN effectiveness, offering novel insights into the social and operational factors that shape crowd-based moderation outcomes.

\section{Methodology}
\subsection{Data} \label{data}
This study draws upon the publicly available daily updated dataset of CNs provided by X. The dataset comprises five principal categories, three of which are utilized in our analysis: (i) Community Notes data, containing the full text of each note, the associated post identifier, and the categorical labels assigned when flagging a post; (ii) Note rating data, which, when aggregated and evaluated in terms of the rater diversity, enables the determination of a note’s perceived helpfulness; (iii) Note status history, documenting the sequence of status changes from initial creation to a permanently locked state after two weeks. 
For our analysis, we retrieved all available records on July 1, 2025, resulting in a corpus of approximately 1.8 million notes. For the research question addressing the time to note success, we defined the observation window with a start date of January 1, 2023, and an end date of May 31, 2025. The selection of this observation window was informed by X’s announcement on January 23, 2023, instituting a policy that locks the status of Community Notes (CNs) two weeks following their creation\footnote{https://x.com/CommunityNotes/status/1616641919173685254}. For notes that had already been active for more than two weeks at the time of the announcement, X applied this locking rule retroactively. To ensure that this change does not affect the natural time to success, we therefore restricted the dataset to the notes created after this policy came into effect. Additionally, we limited the end date to May 31, 2025, to allow sufficient time for each note to reach its final status. This restriction ensures that all notes analyzed for this research question had attained their finite state.

\subsection{How concentrated is the “crowd” in Community Notes?} \label{crowd}
To estimate the extent of concentration in note production among a small group of authors, we conducted an analysis drawing on methods from econometrics, information theory, and competition analysis to quantify ‘dominance’ or ‘inequality’ in contributions. 

First, we calculate a metric of inequality estimation called the \textit{Gini Coefficient} \cite{dorfman1979formula}. Despite being traditionally used in economic welfare domain for income inequality estimation, it can theoretically be applied to a variety of problems, from participation inequality estimation in health-oriented online communities \cite{van2016employing} to citation distribution per author \cite{Petersen_2014}. Since the cumulative number of CNs per user follows a discrete and continuous distribution, the Gini coefficient can be adopted to our task (Equation~\ref{eq:gini}) in order to estimate inequality of contribution among unique authors using.

\begin{equation}\label{eq:gini}
    G = \frac{\sum_{i=1}^{n} \sum_{j=1}^{n} \left| x_i - x_j \right|}{2n^2 \bar{x}}
\end{equation}

where:
\begin{itemize}
    \item $n$ is the total number of individuals (or authors),
    \item $x_i$ is the contribution (e.g., number of notes) of author $i$,
    \item $\bar{x}$ is the mean contribution across all authors.
\end{itemize}

 Gini coefficient ranges from 0 to 1, representing perfect equality and perfect inequality respectively. It is directly derived from the Lorenz curve \cite{gastwirth1971general}, a visual representation of the distribution of income or wealth in a country, illustrating the cumulative share of income held by the cumulative share of the population. 

The second inequality metric we explore is Herfindahl-Hirschman Index (HHI) \cite{rhoades1993herfindahl}, which serves as a statistical measure of concentration. HHI is used to compute concentration in a variety of contexts, including the output of companies in banking or market sectors \cite{Rhoades1993TheHI}. We adapt this metric to our task using Equation~\ref{eq:hhi}. 

\begin{equation}\label{eq:hhi}
HHI = \sum_{i=1}^{n} s_i^2
\end{equation}

where

\begin{itemize}
    \item $s_i = \dfrac{x_i}{\sum_{j=1}^{n} x_j}$ is the share of total activity (number of notes) contributed by author $i$,
    \item $x_i$ is the contribution of author $i$,
    \item and $n$ is the total number of authors.
\end{itemize}

To look into inequality of contribution from the entropy-based perspective, we additionally estimate the Theil Index ($T$) \cite{conceiccao2000young}. $T$ a metric is derived from information theory to quantify the divergence of the observed distribution of contributions from a perfectly equal distribution. 
A Theil Index of $0$ indicates complete equality (all authors contribute equally), while higher values reflect greater inequality, with a small subset of authors producing a disproportionate share of total notes. 

\begin{equation}
T = \frac{1}{N} \sum_{i=1}^{N} \frac{x_i}{\bar{x}} \ln \left( \frac{x_i}{\bar{x}} \right)
\end{equation}

where

\begin{itemize}
    \item $x_i$ is the contribution (e.g., number of notes written) by author $i$,
    \item $\bar{x} = \frac{1}{N} \sum_{i=1}^{N} x_i$ is the mean contribution across all $N$ authors,
    \item and $N$ is the total number of authors.
\end{itemize}

To measure the diversity in note authorship, we estimate Shannon Entropy ($H$) \cite{bromiley2004shannon}. It quantifies the unpredictability of selecting a random contribution's author, with higher entropy indicating 
a more uniform distribution of contributions. Together with the Theil Index, entropy provides a complementary perspective on the concentration and equality of participation.

\begin{equation}
H = - \sum_{i=1}^{N} p_i \ln(p_i)
\end{equation}

where

\begin{itemize}
    \item $p_i = \dfrac{x_i}{\sum_{j=1}^{N} x_j}$ is the proportion of total contributions made by author $i$,
    \item $x_i$ is the number of notes (or total activity) from author $i$,
    \item and $N$ is the total number of authors.
\end{itemize}

Because entropy depends on the number of contributors N, it is often normalized to range between 0 and 1. The normalized form ($H^{*}$) rescales entropy by its theoretical maximum, allowing comparisons across datasets with different numbers of authors. 

\begin{equation}
H^{*} = \frac{H}{\ln(N)}
\end{equation}

Here, $H^*=1$ indicates maximum diversity / perfect equality when all the authors contribute equally, while $H^*=0$ indicates complete concentration (a single author produces all contributions).

Finally, we calculate the top-$k$ share ($S_k$) to measure the proportion of total contributions made by the most active $k$ authors. For instance, a Top-1\% Share of $0.28$ indicates that the top 1\% of contributors produce 28\% of all notes. 
This metric provides an intuitive view of concentration that complements formal inequality indices such as the Gini coefficient or Theil index. Higher values of $S_k$ signify stronger dominance by a small elite of contributors. In our analysis, we focused on top-1\% and top-10\% measures.

For each of the metrics described above, we additionally explore how concentration (inequality) changes over time. This provides further insight into whether the note-writing activity is becoming more dominated by a few people during certain periods of time. To perform such an analysis, we used the timestamp of note creation to estimate the monthly contributions per note author. We then computed the inequality metrics over the monthly period of the CN dataset.

\subsection{Fostering consensus or dissensus?}
To further analyse the types of engagement produced by the crowd, we examine the patterns of consensus building. The extent of the consensus building through CNs can be operationalised through two metrics. First, for the posts that received more than one CN, the notes can agree or disagree on whether a particular post is \textit{misinformation} or \textit{not misinformation}. We refer to this dimension as ``classification consensus''. The second method of defining consensus is based on the agreement of the crowd regarding the helpfulness of each individual note. Under this definition, consensus may be reached for both published and unpublished notes. We refer to it as ``rating consensus''. 

There are three potential outcomes for the \textit{classification consensus}, and each concerns posts receiving more than one note. 
\begin{itemize}
\item \textit{Dissensus}: There is no agreement among the note posters on whether a post on X is misinformation or not misinformation. By the lack of agreement, we mean that there is a mixture of notes marking the post as misinformation or not.
\item \textit{Consensus --- not misleading}: All the notes that are created for a particular post mark it as not misinformation.
\item \textit{Consensus --- misleading}: All the CNs that are flagging the post marked it as misinformation.
\end{itemize}

To estimate the extent of classification consensus, we analysed the temporal distribution of consensus and dissensus by grouping the notes related to the same unique post and the same ``classification'' field.

For the rating consensus, we define consensus as the presence of agreement on whether the note should be published. To estimate the extent of the rating consensus, we selected the notes within the ``Note Status History'' dataset that have the status of ``needs more ratings'' (NMR). As per X's definition, this status serves as an indicator that the note has not yet met the algorithmic consensus criterion of a “diversity of perspectives” and therefore has not been agreed upon as either helpful or unhelpful. 

While the CN system is designed to render a judgement, dissensus may occur potentially owing to user misuse, trolling behaviour, or strategic non-cooperation. Thus, dissensus is not necessarily organic. In certain cases, it may be deliberately evoked, promoted, or sustained as part of ongoing contestation over a note’s validity.

One of such misuse cases is the NNN phenomenon described in Section~\ref{related:platform}. Rather than flagging potentially misleading content, participants use the platform to engage in arguments by providing counterpoints and meta-commentary.

To examine the role of Note Not Needed (NNN) in shaping patterns of dissensus, we conducted a complementary analysis by identifying all posts that either contain the term “NNN” or “Note Not Needed” in their note text, or have at least one rating marking at least one of its notes as not helpful for the reason “Note Not Needed”. We then examine the temporal distribution of this engagement, focusing on posts associated with notes flagged as NNN to assess how the crowd misuses the system over time.

We also performed a statistical analysis using Odds Ratios (OR) to assess whether the \textit{Note Not Needed} (NNN) phenomenon is associated with the level of helpfulness consensus among notes attached to a given post \cite{mchugh2009odds}. Our assumption is that if at least one person indicated that the post needs no notes, we can assume that the post represents an opinion rather than a false claim. One would therefore expect the successful crowdsourcing community to agree on non-helpfulness of any notes under such posts. Under our analysis, we define \textit{NNN posts} as posts that have at least one note flagged as NNN, either through the note taking or note rating crowd engagement. Similarly, a \textit{non-NNN posts} are the posts that had no notes flagged as NNN. This leads to the following $2\times 2$ contingency table:
\[
\begin{array}{c|cc}
 & \text{Published (True)} & \text{Not Published (False)} \\
\hline
\text{NNN-flagged (True)} & a & b \\
\text{Not NNN (False)} & c & d \\
\end{array}
\]
where:
\begin{itemize}
    \item $a$ = number of NNN posts that contain at least one published note,
    \item $b$ = number of NNN posts that have no any published notes,
    \item $c$ = number of non-NNN posts that have at least one published note,
    \item $d$ = number of non-NNN posts that have no NNN-flagged notes and that have no notes published.
\end{itemize}

We then test whether the NNN posts are less likely to have any notes published compared to non-NNN posts under the following hypotheses:

\[
\begin{aligned}
H_0 &: \; P(\text{Published} \mid \text{NNN}) = P(\text{Published} \mid \text{Not NNN}) \\
H_1 &: \; P(\text{Published} \mid \text{NNN}) < P(\text{Published} \mid \text{Not NNN})
\end{aligned}
\]

Based on these hypotheses, we estimate OR as a quantifier of the association between NNN flagging and people refusing to accept the notes under such posts:

\begin{equation}
\text{OR} = \frac{a/b}{c/d} = \frac{ad}{bc}
\end{equation}

According to the OR interpretation, $\text{OR} < 1$ indicates that NNN-flagged notes are less likely to be published compared to other notes.

\subsection{Is consensus more likely the outcome of the timeliness of notes?} \label{sec:crh}

As discussed in Section~\ref{sec:effectiveness}, prior research has emphasized that the timeliness of notes plays a critical role: the inherently slower pace of crowdsourced moderation can limit its effectiveness in preventing the spread of misinformation before it gains traction. Because the CN bridging algorithm requires a sufficient diversity of perspectives for a note to become visible, the time needed to meet this threshold may extend beyond the period in which most misinformation is spread. We therefore aim to examine more closely whether the timeliness of a note itself can enhance the effectiveness of consensus building.

To answer this question, we first retrieve the exact time of each post associated with a community note data using X’s Snowflake library for Python\footnote{https://github.com/twitter-archive/snowflake}. We then calculate the time lag between the post on X and note publication time, subtracting the timestamp of the post from the earliest note publication timestamp. To calculate the lag, we utilize the Note Status History dataset, which records several relevant timestamps:

\begin{itemize}
    \item The first instance when the note’s status transitioned from NMR to either “Helpful” or “Not Helpful”.
    \item The timestamp corresponding to the change to the current status (either “Helpful” or “Not Helpful”).
    \item The time of the most recent non-NMR status update.
    \item The time when the note was locked, i.e., when its status became permanent as either “Helpful”, NMR or “Not Helpful”.
\end{itemize}

To identify this earliest time, we utilize a two-step filtering procedure:
\begin{itemize}
    \item Eligibility check: Filter in only the notes with a locked status of Currently Rated Helpful or Currently Rated Not Helpful.
    \item Earliest status timestamp: For each eligible note, search all status transitions to identify the first moment the note entered its final state.
\end{itemize}

We then group the calculated time lags into 3-hour intervals and stratify them by note status to estimate the temporal window during which note creation is most likely to maximize publication likelihood.

\section{Results}

\subsection{Crowd composition analysis}\label{result:crowd}

We examined whether note creation is broadly distributed across contributors or is concentrated among a small subset of highly active authors. To quantify this, we calculated multiple inequality measures — the Gini coefficient, Herfindahl–Hirschman Index (HHI), Theil index, and normalized Shannon entropy — alongside the cumulative contribution shares of the top 1\%, 10\% of authors. The results of the cumulative analysis are provided in Table~\ref{tab:crowd}.

\begin{table}[h] 
\centering
\caption{Summary of inequality metrics for note authorship distribution.}
\label{tab:inequality_metrics}
\begin{tabular}{l c}
\hline
\textbf{Metric} & \textbf{Value} \\
\hline
Gini coefficient & 0.679 \\
Herfindahl--Hirschman Index (HHI) & 0.0009 \\
Theil index & 1.442 \\
Normalized entropy & 0.884 \\
Top 1\% share & 0.282 \\
Top 10\% share & 0.584 \\
\hline
\end{tabular}
\label{tab:crowd}
\end{table}

\begin{figure*}[ht!]
    \centering
    \includegraphics[width=0.8\linewidth]{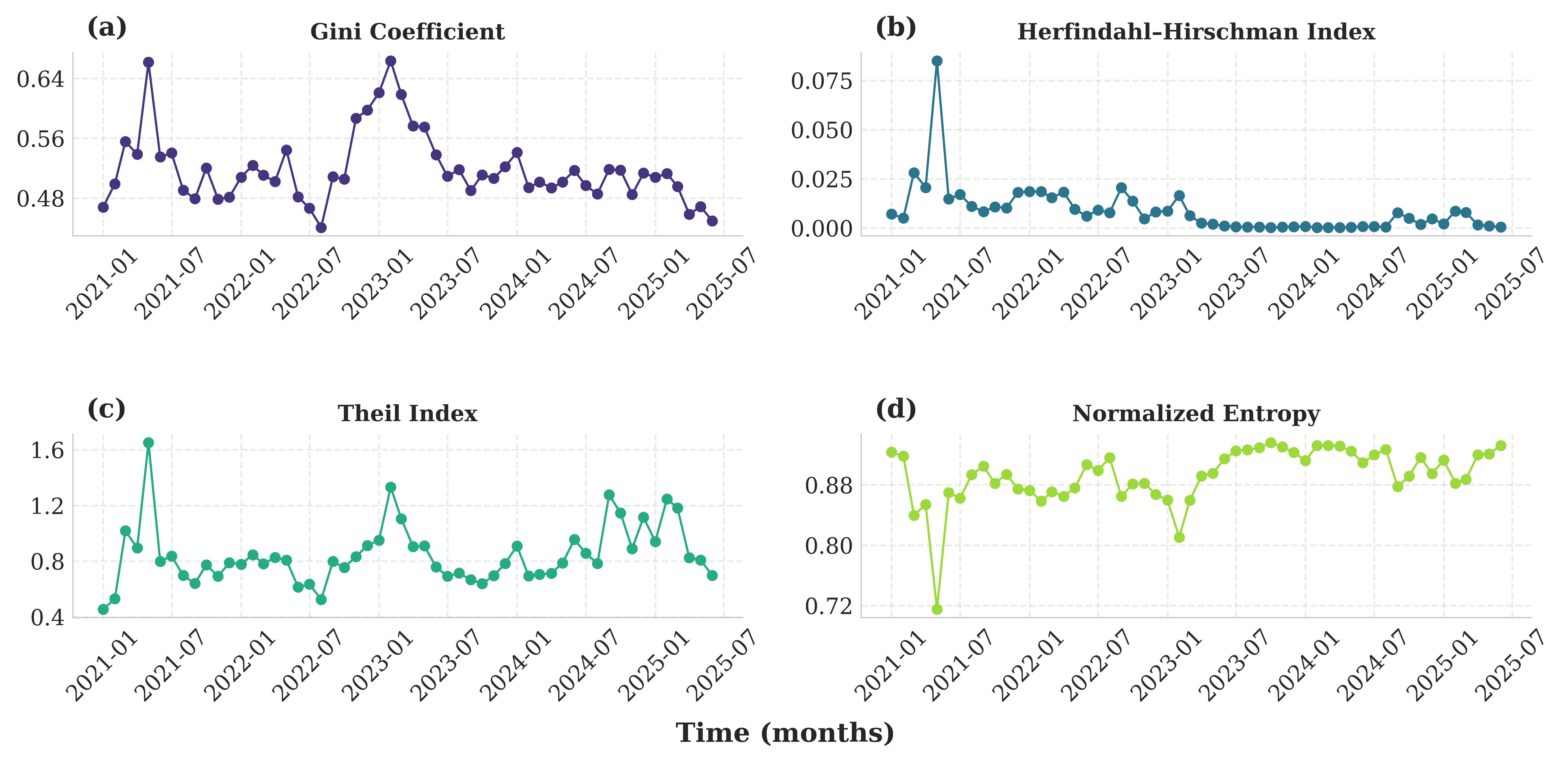}
    \caption{Inequality of Note Contributions Over Time.}
    \label{fig:temporal}
\end{figure*}

Overall, our analysis reveals that note authorship is strongly skewed, with a Gini coefficient $\approx0.68$, indicating that contributions are far from evenly distributed. In many social systems or collaborative content platforms, Gini values in the 0.5–0.8 range are consistent with “heavy-tailed” contribution patterns (i.e., a few contributors are responsible for a large share of activity). For instance, in studies of scientific publishing, Gini values of 0.73 for citation counts reflect extreme skew in impact among authors \cite{Petersen_2014}. 
The observed value of $\approx0.68$ for CN data therefore indicates a high degree of inequality consistent with a platform where a more committed core of users dominates contribution.

Although the HHI metric is numerically small ($\approx0.000878$), in a system with many authors, even small deviations from uniformity imply meaningful concentration. Such levels of inequality are comparable to skew observed in other high-activity platforms \cite{van2016employing}.

While normalized entropy ($\approx0.88$) suggests that the system maintains a broad base of participants, the top 1\% of authors account for $\approx$28\% of all notes, and the top 10\% comprise $\approx$58\%, underscoring the dominance of a relatively small elite. In cumulative participation literature, such “skewed top-share” patterns are well-known: for instance, in health-oriented platforms or scientific citations, a small fraction of participants or scholars account for a disproportionately large fractions of impact \cite{Petersen_2014, van2016employing}.

The Theil index ($\approx1.44$) further confirms substantial divergence from an ideal uniform distribution. This pattern suggests that the system’s productivity is centralized: while many users contribute, a few exert disproportionate influence.

The temporal analysis of the concentration metrics is shown in Figure~\ref{fig:temporal}.
It reveals distinct evolutionary phases. During 2021–2022, participation was highly concentrated, with Gini values exceeding 0.6 and Theil indices above 1.5, reflecting dominance by a small cohort of early contributors. As the platform expanded globally through 2023, inequality initially intensified — consistent with cumulative advantage dynamics observed in open collaboration systems \cite{doi:10.1126/science.286.5439.509, crowston2012floss}. However, from late 2023 onward, the Gini coefficient declined to $\approx0.5$ and normalized entropy rose above 0.93, indicating increasing diversification of contributors. This suggests that the system has matured toward a more balanced state. Such evolution aligns with participation cycle theories \cite{panciera2009wikipedians}, where initial concentration transitions to broader engagement as communities scale. Overall, the results depict a platform moving from elite-dominated origins toward a more inclusive participation regime, while retaining residual concentration characteristic of online moderation ecosystems.

\subsection{Do CNs tend to build consensus?}

\begin{figure}[ht!]
    \centering
    \includegraphics[width=1\linewidth]{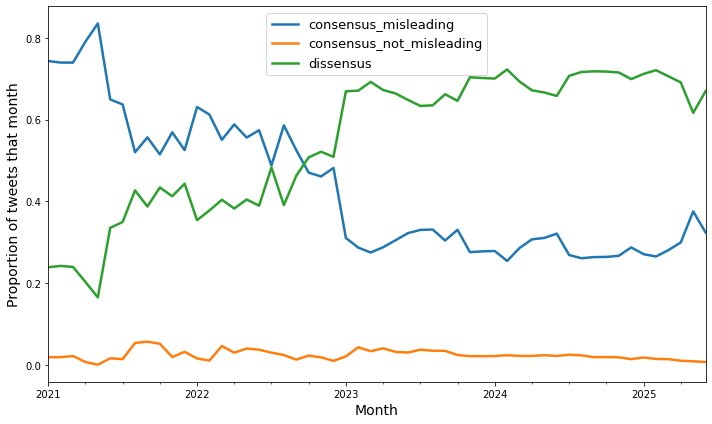}
    \caption{Note classification agreement per month.}
    \label{fig:dissensus}
\end{figure}

Our analysis suggests that CNs are ineffective in building consensus. In particular, we found that the prevalent proportion of notes ($\approx69\%$) are in a dissensus about the post being misleading or not, with only $\approx29\%$ of notes having a consensus regarding a potentially misleading post and a small proportion of notes ($\approx2\%$) in a  consensus regarding a post being not misinformation. The temporal analysis (Figure~\ref{fig:dissensus}) further reveals a strong trend towards the prevalence of dissensus over time as more people author the notes. While the early system, with its few note authors, produced a consensus about a substantial majority of posts being misleading, with the global enrolment of CNs in early 2023, the proportion of cases of disagreement increased to 69\%. Given the change in crowd diversification discussed in Section~\ref{result:crowd}, this finding can be attributed to a variety of opinions and a higher percentage of active contributors.

With regard to the rating consensus, we found that raters reached consensus on 11.5\% of all notes, with the remaining 88.5\% of the notes remaining in the NMR status. This could be attributed to two factors. First, some of the notes never receive 5 ratings to move out of the initial NMR state. Second, the helpfulness estimation model may never reach the required diversity of perspective in the ratings, thereby not satisfying the algorithmic criteria.

\begin{figure*}
    \centering
\includegraphics[width=0.8\linewidth]{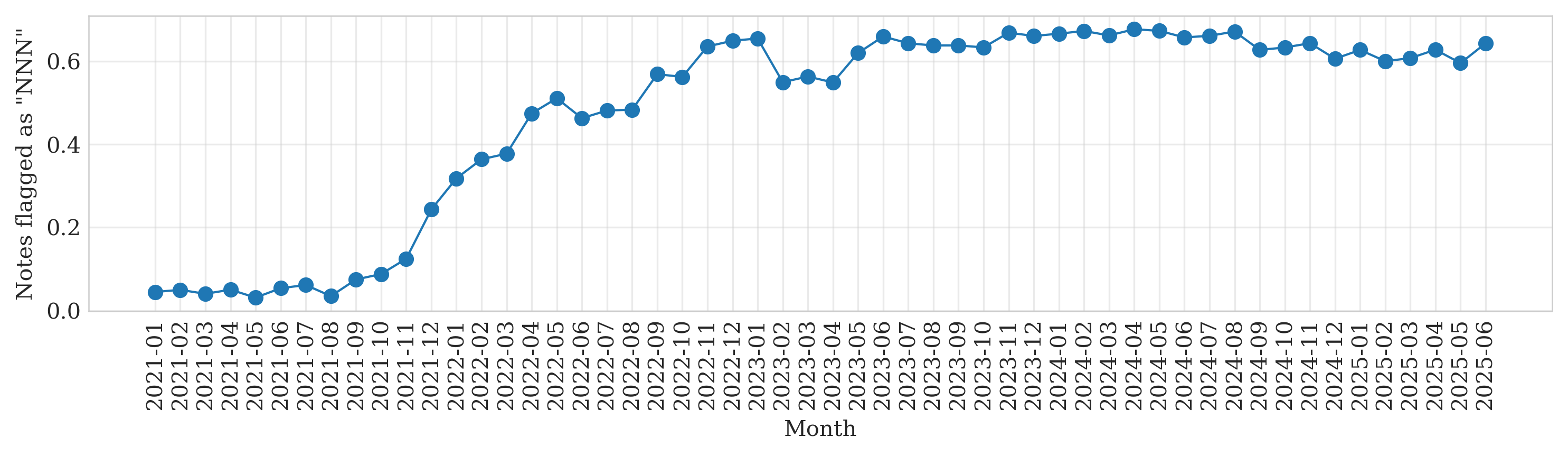}
    \caption{Monthly proportion of posts flagged as NNN.}
    \label{fig:nnn}
\end{figure*}

\textbf{The “NNN” phenomenon.} The distribution of the post proportion that exhibits the NNN phenomenon is illustrated in Figure~\ref{fig:nnn}. As can be seen, the prevalence in dissensus notes reported in Figure~\ref{fig:dissensus} is aligned with a spike in the NNN-noted posts toward the end of 2022--beginning of 2023, suggesting that this phenomenon can be one driver of dissensus. After this period, the proportion of NNN-noted posts stabilises at a high level of $\approx68\%$ of all posts. This observation highlights the persistence of what we term``noting as debating'', a dynamic that can seen as a form of system repurposing, where participants use a tool intended for content evaluation as a forum for interaction and contestation. While this constitutes a “misuse” of the system from the perspective of platform design and policy, it may also indicate an unmet user demand for structured discussion spaces within the platform ecosystem.  
\begin{table}[h!]
\centering
\caption{Contingency table showing the relationship between NNN presence and note publication status.}
\begin{tabular}{lcc}
\toprule
\textbf{NNN Presence} & \textbf{No Published Note} & \textbf{Has Published Note} \\
\midrule
Absent & 456{,}727 & 23{,}599 \\
Present & 684{,}282 & 110{,}461 \\
\bottomrule
\end{tabular}
\label{tab:contingency}
\end{table}

Contrary to our expectation, we found no significant association between the NNN posts and decreased likelihood of publication of any notes under that post, with an Odds Ratio$ = 3.124$ and $p<10^{-300}$. Table~\ref{tab:contingency} represents the contingency table for Fisher's exact test. This finding indicates that the use of Community Notes for debate or opinion-driven exchanges is not merely incidental but appears to be reinforced by community feedback, as such notes are disproportionately marked as helpful.

\subsection{The timeliness of CN posting}

\begin{figure*}
    \centering
    \includegraphics[width=0.55\linewidth]{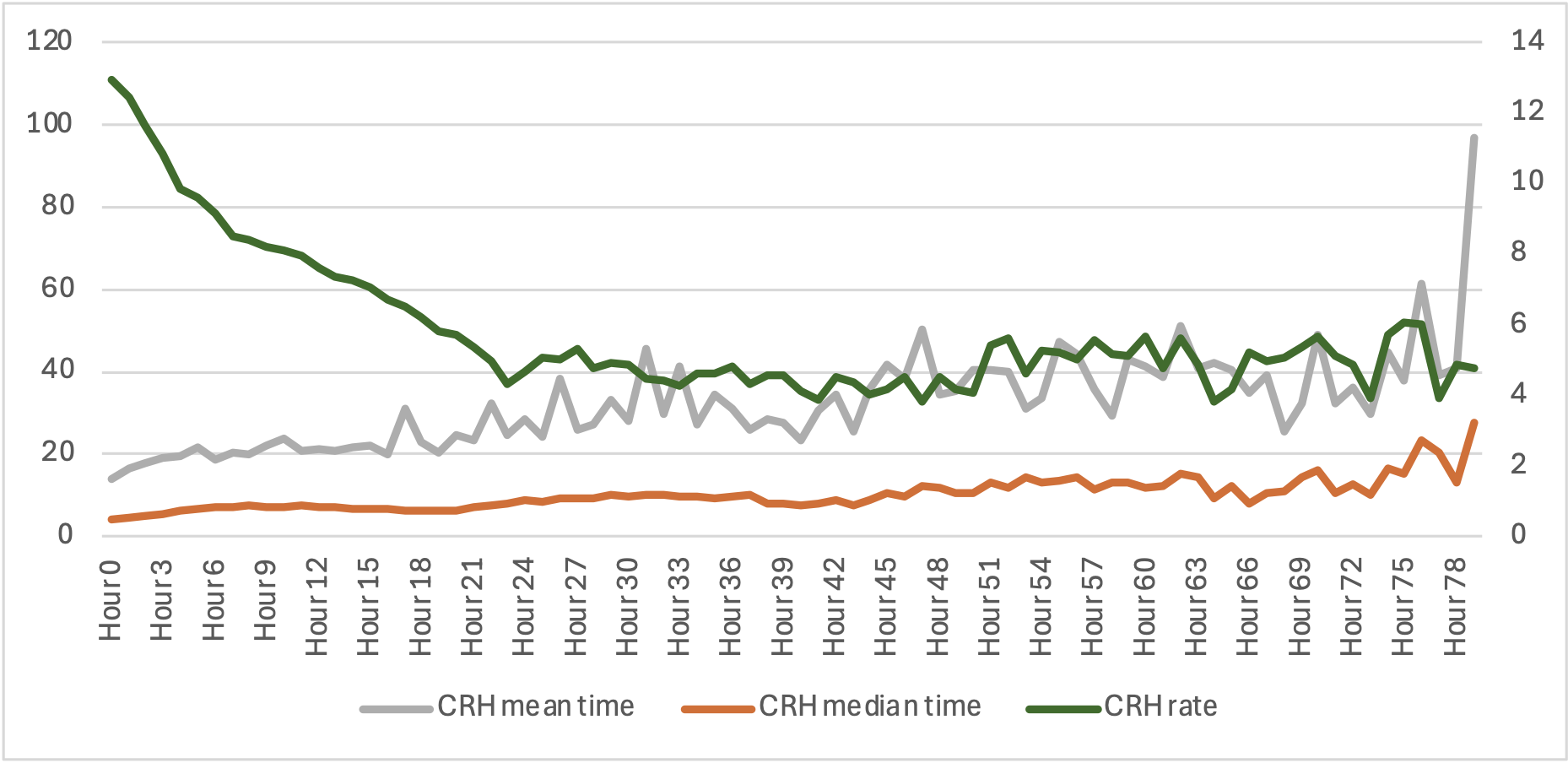}

    \caption{Effect of Creation Delay on Ratio and Time to Reach CRH in the First 80 Hours}
    \label{fig:4}
\end{figure*}

Figure~\ref{fig:4} illustrates the success rate of Community Notes (CRH) in the context of the timeliness of writing a CN after the publication of the post on X. CRH exhibits dependency on the delay in note creation relative to the time of the post, which is associated with both longer times and lower percentages of reaching to CRH status. We calculated Spearman’s rank correlation to assess the monotonic relationships between the creation delay and two outcome variables: mean time to CRH and percentage of CRH. The analysis reveals a statistically significant positive monotonic relationship between the creation delay and the mean time to CRH (rs=0.8633, p<0.001), indicating that longer delays are associated with increased time to become helpful. In contrast, we observed a significant negative monotonic relationship between the creation delay and the percentage of helpful notes (rs=-0.5118, p<0.001), showing that as the delays increase, the proportion of helpful notes decreases. These findings suggest that longer creation delays negatively impact the timeliness and helpfulness of notes.

On average, notes reach a \textit{Currently Rated Helpful (CRH)} status 21.7 hours after creation, with a median time of 6.2 hours. In comparison, notes classified as \textit{Currently Rated Not Helpful (CRNH)} take longer to stabilize, averaging 43.7 hours with a median of 5.0 hours. When considering the full timeline from post publication to note evaluation, the temporal gap becomes more pronounced. Posts receive their first helpful note after an average of 65.7 hours ($\approx2.7$ days) and a median of 15.3 hours, while notes ultimately deemed not helpful appeared later—after an average of 94.3 hours ($\approx3.9$ days) and a median of 18.6 hours. These findings suggest that helpful notes tend to emerge more quickly, whereas unhelpful notes are both slower to appear and slower to converge on their final status.
\section{Discussion}
Our findings highlight a fundamental tension between the design aspirations of CNs and their empirical outcomes. The platform aims to crowdsource content moderation by harnessing diversity to reach consensus; however, our analyses show that this diversity often amplifies disagreement instead. In particular, we observed a correlation between the time when the concentration of the crowd became less skewed and the time when disagreement among the note writers became prevalent, with nearly 69\% of notes being in disagreement with each other regarding the presence of misleading content. Our analysis of the note moderation consensus also revealed a high state of disagreement, with 90\% of notes remaining unresolved. This persistent dissensus may stem from either an insufficient number of contributor ratings or a lack of the “diverse perspectives” required by the platform’s bridging algorithm. 

A key factor underlying this dissensus is the widespread occurrence of the notes under the posts that ``need no notes'' (NNN posts). We observed not only the persistent prevalence of notes attached to opinion-based posts but also that such notes frequently achieve successful publication. Several factors may explain this pattern. First, such posts may potentially combine factual claims with subjective commentary, leading contributors to disagree on whether they warrant annotation. In such cases, some notes may legitimately address misleading elements, while others invoke the NNN label to signal that opinionated content requires no correction. Second, some notes may have been created and published before subsequent contributors marked the post as NNN. Finally, highly controversial or widely circulated opinionated posts attract more overall engagement, increasing the likelihood that at least one note will meet the publication criteria despite the underlying disagreement.

Timeliness further constrains CN effectiveness. On average, notes that are timely in terms of combating potentially misleading content are significantly more likely to be published. However, we still observed an over 65 hour delay in publishing the note after the emergence of a misleading post. Therefore, contrary to the belief that community-based moderation offers a more timely reaction to misinformation, this delay is comparable to that of professional fact-checkers  that is reported to take up to four days on average \cite{Wack2024PoliticalFactChecking, Micallef2022TrueFalse}. As a result, CNs primarily function as post-hoc correctives rather than preventative interventions.

Finally, participation inequality remains a defining characteristic of CNs.  We found that a small minority (approximately 10\%) of contributors account for 58\% of all notes, while the majority remain largely inactive. Such concentration mirrors the epistemic asymmetries seen in traditional fact-checking, centralizing influence in a narrow elite. If unchecked, this concentration may limit diversity of perspectives, increase fragility to contributor churn, and entrench power among the most active authors. Our temporal analysis suggests that the platform seems to be slowly moving toward the more diverse and truly crowdsource-based state. 

Together, these results challenge the foundational assumption of CNs as a form of effective collective moderation. Instead of aggregating distributed expertise, CNs operate as a stratified and temporally constrained system shaped by a small subset of users—an algorithmic crowd whose diversity is more formal than substantive.

\section{Conclusions and Future Work}

This study demonstrates that Community Notes, while designed to democratize moderation, often reproduce the very asymmetries they were meant to overcome. The system’s reliance on algorithmic diversity produces persistent dissensus, its timeliness fails to match the velocity of misinformation, and its participation dynamics concentrate epistemic authority within a small core of contributors. 

To realise the democratic and epistemic potential of crowdsourced moderation, future designs must integrate deliberative affordances— structured opportunities for transparent dialogue— alongside algorithmic scaling. Introducing features that separate fact annotation from meta-discussion could preserve CNs’ corrective value while accommodating legitimate debate. Additionally, quality-weighted diversity metrics and balanced participation incentives could mitigate contributor inequality. Going forward, decomposing inequality across content categories and tracking its evolution longitudinally will be critical to assessing whether the platform is becoming increasingly concentrated and to inform interventions aimed at broadening participation.

Future research should examine the longitudinal evolution of Community Notes participation, particularly during major political events, to better understand how patterns of dissensus and contribution concentration shift over time. Moreover, persistent asymmetries in note production raise critical questions about the potential for coordinated or strategic behaviours, including brigade-style interventions in both note writing and rating \cite{prollochs_community-based_2022, chuai_community_2025}. Investigating these phenomena through network or temporal anomaly detection methods could help distinguish organic crowd participation from orchestrated collective action. Such analyses would advance our understanding of the social and algorithmic mechanisms underlying crowdsourced moderation, offering a foundation for more resilient, transparent, and democratically accountable systems.

\begin{acks}

This  work  has  been  co-funded  by  the  UK’s innovation agency (Innovate UK) grant 10039055 (approved under the Horizon Europe Programme as vera.ai,  EU  grant  agreement  101070093) under action number 2020-EU-IA-0282. The contribution by Richard Rogers is funded by the Horizon Europe Programme project, SoMe4Dem, grant nr. 101094752. We would like to thank Bumju Jung, Maricarmen Rodríguez Guillen, and Wilma Ewerhart for their contributions to this research during 2025 Digital Methods Summer School, Media Studies, University of Amsterdam. The contribution by Adel Tayebi is financed by CY Cergy Paris University as part of the EUTOPIA co-tutelle scheme.
\end{acks}

\bibliographystyle{ACM-Reference-Format}
\bibliography{sample-base}

\appendix








\end{document}